\documentclass[11pt, a4paper]{article}
\usepackage{latexsym}
\usepackage{graphicx}
\usepackage{cite}
\usepackage{amsmath}
\usepackage{amssymb}

\input{tcilatex}

\begin{document}

\title{Investigating Diffusion Coefficient Using Dynamic Light Scattering
Technique}
\author{Yong Sun{$\thanks{%
Email: ysun2002h@yahoo.com.cn}$}}
\maketitle

\begin{abstract}
In this work, the Z-average, effective, apparent diffusion coefficients and
their poly-dispersity indexes were investigated for dilute poly-disperse
homogeneous spherical particles in dispersion where the Rayleigh-Gans-Debye
approximation is valid. The results reveal that the values of the apparent
and effective diffusion coefficients at a scattering angle investigated are
consistent and the difference between the effective and Z-average diffusion
coefficients is a function of the mean particle size, size distribution and
scattering angle. For the small particles with narrow size distributions,
the Z-average diffusion coefficient can be got directly at any scattering
angle. For the small particles with wide size distributions, the Z-average
diffusion coefficient should be measured at a small scattering angle. For
large particles, in order to obtain a good approximate value of Z-average
diffusion coefficient, the wider the particle size distribution, the smaller
the scattering angle that the DLS data are measured. The poly-dispersity
index of the effective diffusion coefficient at a scattering angle
investigated is consistent with that of the Z-average diffusion coefficient
and without considering the influences of noises, the difference between the
poly-dispersity indexes of the Z-average and apparent diffusion coefficients
is determined by the mean particle size, size distribution and scattering
angle together.
\end{abstract}

\section{INTRODUCTION}

For colloidal dispersion systems, light scattering is a widely used
technique to measure the characteristics of particles. One of the main
applications of the dynamic light scattering (DLS) technique is to obtain
the Z-average diffusion coefficient and its poly-dispersity index for
particles in liquid suspension. The DLS technique is to measure the particle
properties from the normalized time auto-correlation function of the
scattered light $g^{\left( 2\right) }\left( \tau \right)$, here $\tau$ is
the correlation delay time. The moments (or Cumulants) method \cite%
{re1,re2,re3,re4} has been used as a standard method to measure the
Z-average diffusion coefficient and its poly-dispersity index from the DLS
data. In general, when the DLS data are analyzed, the mean diffusion
coefficient and its poly-dispersity index obtained using the moments method
are the apparent diffusion coefficient and its poly-dispersity index.

About the relationship between the Z-average and apparent diffusion
coefficients has been investigated by a few authors\cite{re3}. Under
appropriate conditions, they measured the Z-average diffusion coefficient by
fitting $ln\left( C^{1/2}\left\vert g^{\left( 1\right) }\left( \tau \right)
\right\vert \right) $ to linear, quadratic and cubic function of $\tau$. The
effective values of $\langle \Gamma\rangle$ and $\mu_2$ obtained from the
fits are plotted against $\langle \Gamma\rangle \tau_{max}$ and thus are
obtained by extrapolating $\langle \Gamma\rangle \tau_{max}$ to 0, where $%
\langle \Gamma\rangle$ is the mean decay rate. Combining in the Svedberg
equation with the weight-average sedimentation coefficient, the
weight-average molar mass can be yielded. For large particles, the authors
suggest the Z-average diffusion coefficient should be measured at a small
enough scattering angle. In order to obtain the accurate value of the
weight-average molar mass, it is important to measure the value of Z-average
diffusion coefficient accurately.

In this work, the Z-average, effective, apparent diffusion coefficients and
their poly-dispersity indexes were investigated for dilute poly-disperse
homogeneous spherical particles in dispersion where the Rayleigh-Gans-Debye
(RGD) approximation is valid. The results reveal that the values of the
apparent and effective diffusion coefficients at a scattering angle
investigated are consistent and the difference between the effective and
Z-average diffusion coefficients is a function of the mean particle size,
size distribution and scattering angle. For the small particles with narrow
size distributions, the Z-average diffusion coefficient can be got directly
at any scattering angle. For the small particles with wide size
distributions, the Z-average diffusion coefficient should be measured at a
small scattering angle. For large particles, in order to obtain a good
approximate value of Z-average diffusion coefficient, the wider the particle
size distribution, the smaller the scattering angle that the DLS data are
measured. The poly-dispersity index of the effective diffusion coefficient
at a scattering angle investigated is consistent with that of the Z-average
diffusion coefficient and without considering the influences of noises, the
difference between the poly-dispersity indexes of the Z-average and apparent
diffusion coefficients is determined by the mean particle size, size
distribution and scattering angle together. For narrow particle size
distributions, they are consistent. For wide particle size distributions,
the poly-dispersity index of apparent diffusion coefficient should be
measured at a $\tau $ range where $\tau _{max}$ changes to a small value in
order to obtain a good approximate result of the poly-dispersity index of
Z-average diffusion coefficient.

\section{THEORY}

For dilute poly-disperse homogeneous spherical particles in dispersion where
the RGD approximation is valid, the effective diffusion coefficient $D_{eff}$
and its poly-dispersity index $\left(\frac{\sqrt{\mu _{2}}}{\langle
\Gamma\rangle}\right)_{eff}$ obtained using the moments method at a given
scattering vector as $\tau\rightarrow0 $ are given by

\begin{equation}
D_{eff}=\frac{\int_{0}^{\infty }R_{s}^{6}P\left( qR_{s}\right) DG\left(
R_{s}\right) dR_{s}}{\int_{0}^{\infty }R_{s}^{6}P\left( qR_{s}\right)
G\left( R_{s}\right) dR_{s}}  \label{Deff}
\end{equation}

\noindent and

\begin{equation}
\left(\frac{\sqrt{\mu _{2}}}{\langle \Gamma\rangle}\right)_{eff}=\left( 
\frac{\int_{0}^{\infty }R_{s}^{6}P\left( qR_{s}\right) D^{2}G\left(
R_{s}\right) dR_{s}\int_{0}^{\infty }R_{s}^{6}P\left( qR_{s}\right) G\left(
R_{s}\right) dR_{s}}{\left( \int_{0}^{\infty }R_{s}^{6}P\left(
qR_{s}\right)D G\left( R_{s}\right) dR_{s}\right) ^{2}}-1\right) ^{1/2}
\label{Deffindex}
\end{equation}
where $R_{s}$ is the static radius, $D$ is the diffusion coefficient, $\ q=%
\frac{4\pi }{\lambda }n_{s}\sin \frac{\theta }{2}$ is the scattering vector, 
$\lambda $\ is the wavelength of the incident light in vacuo, $n_{s}$ is the
solvent refractive index, $\theta $ is the scattering angle, $\mu _2$ is the
second moment, $G\left( R_{s}\right) $ is the number distribution of
particles and the form factor $P\left( q,R_{s}\right) $ is

\begin{equation}
P\left( q,R_{s}\right) =\frac{9}{q^{6}R_{s}^{6}}\left( \sin \left(
qR_{s}\right) -qR_{s}\cos \left( qR_{s}\right) \right) ^{2}.  \label{factor}
\end{equation}

If the scattering vector approximates 0, the Z-average diffusion coefficient 
$D_z$ and its poly-dispersity index $\left(\frac{\sqrt{\mu _{2}}}{\langle
\Gamma\rangle}\right)_z$ can be obtained

\begin{equation}
D_z=\frac{\int_{0}^{\infty }R_{s}^{6}DG\left( R_{s}\right) dR_{s}}{%
\int_{0}^{\infty }R_{s}^{6} G\left( R_{s}\right) dR_{s}}  \label{Dzff}
\end{equation}

\noindent and

\begin{equation}
\left(\frac{\sqrt{\mu _{2}}}{\langle \Gamma\rangle}\right)_z=\left( \frac{%
\int_{0}^{\infty }R_{s}^{6}D^{2}G\left( R_{s}\right) dR_{s}\int_{0}^{\infty
}R_{s}^{6}G\left( R_{s}\right) dR_{s}}{\left( \int_{0}^{\infty }R_{s}^{6} D
G\left( R_{s}\right) dR_{s}\right) ^{2}}-1\right) ^{1/2}.  \label{Dzindex}
\end{equation}

\noindent From the Einstein-Stokes relation, the diffusion coefficient $D$
can be written as

\begin{equation}
D=\frac{k_{B}T}{6\pi \eta _{0}R_{h}},
\end{equation}
where $\eta _{0}$, $k_{B}$, $T$ and $R_{h}$ are the viscosity of the
solvent, Boltzmann's constant, absolute temperature and hydrodynamic radius,
respectively.

In this work, the number distribution is chosen as a Gaussian distribution

\begin{equation}
G\left( R_{s};\left\langle R_{s}\right\rangle ,\sigma \right) =\frac{1}{
\sigma \sqrt{2\pi }}\exp \left( -\frac{1}{2}\left( \frac{R_{s}-\left\langle
R_{s}\right\rangle }{\sigma }\right) ^{2}\right) ,
\end{equation}
where $\left\langle R_{s}\right\rangle $ is the mean static radius and $%
\sigma $ is the standard deviation related to the mean static radius.

\section{RESULTS AND DISCUSSION}

In the previous work\cite{re5,re6}, it was shown that the expected values of
the DLS data calculated based on the commercial and static particle size
information are consistent with the experimental data. In order to
investigate the quantities of $D_z$, $D_{eff}$, $D_{app}$ and their
ploy-dispersity indexes accurately, the values of $g^{\left( 2\right)
}\left( \tau \right)$ were produced as described in the previous work\cite%
{re5} and thus $D_z$, $D_{eff}$ and their ploy-dispersity indexes are
calculated using Eqs.\ref{Deff}, \ref{Deffindex}, \ref{Dzff}, \ref{Dzindex}
and $D_{app}$ and its poly-dispersity index are obtained using the first two
moments method, respectively.

The values of $g^{\left( 2\right) }\left( \tau \right) $ were produced using
the information: the temperature $T$, viscosity of the solvent $\eta _{0}$,
wavelength of laser light $\lambda $, refractive index of the water $n_{s}$
and constant $R_{h}/R_{s}$ were set to 300.49K, 0.8479 mPa$\cdot $S, 632.8
nm, 1.332 and 1.1 for $\left\langle R_{s}\right\rangle =$50 nm, 1.2 for $%
\left\langle R_{s}\right\rangle =$120 and 260 nm, scattering angle $\theta $
was chosen as 30$^{\text{o}}$ and 90$^{\text{o}}$ for $\left\langle
R_{s}\right\rangle =$50 and 120 nm and 30$^{\text{o}}$ for $\left\langle
R_{s}\right\rangle =$260 nm, and standard deviation $\sigma $ was set to (3,
5, 10, 15, 20, 25)nm, (8, 12, 24, 36, 48, 60)nm and (16, 26, 52, 78, 104,
130)nm for $\left\langle R_{s}\right\rangle =$50, 120 and 260 nm,
respectively. When the data of $\left( g^{\left( 2\right) }\left( \tau
\right) -1\right) /\beta $ were obtained, the random errors were set 3\%. In
order to investigate the poly-dispersity indexes for narrow particle size
distributions, the statistical noises were not added.

When the first two moments method were used to fit the simulated data, the
values of $\Gamma _{app}$ and $\mu _{2app}$ were chosen when they stabilize.
The results for the simulated data produced based on the mean static radius
50 nm and standard deviations 5, 25 nm at scattering angles 30$^{\text{o}}$
and 90$^{\text{o}}$ were shown in Figs. 1 and 2, respectively. The results
show that all the fit values obtained using the first two moments method are
consistent with the simulated data very well, respectively.

The values of $D_{z}$ and $D_{eff}$ at scattering angles 30$^{\text{o}}$ and
90$^{\text{o}}$ for different standard deviations 3, 5, 10, 15, 20 and 25 nm
were calculated using Eqs. \ref{Deff} and \ref{Dzff} and the results of $%
D_{app}$ are obtained using the first two moments method to fit the
different simulated data, respectively. All results of $D_{z}$, $D_{eff}$
and $D_{app}$ for standard deviations 3, 15 and 25 nm are listed in Table %
\ref{50Dall}.

\begin{table}[ht]
\begin{center}
\begin{tabular}{|c|c|c|c|c|c|c|}
\hline
$\sigma \left( nm\right) $ & \multicolumn{2}{|c|}{3} & \multicolumn{2}{|c|}{
15} & \multicolumn{2}{|c|}{25} \\ \hline
$D_{z}\left(10^{-12}m^2/s\right)$ & \multicolumn{2}{|c|}{4.638} & 
\multicolumn{2}{|c|}{3.501} & \multicolumn{2}{|c|}{2.687} \\ \hline
$\theta $ & 30$^\mathrm o$ & 90$^\mathrm o$ & 30$^\mathrm o$ & 90$^\mathrm o$
& 30$^\mathrm o$ & 90$^\mathrm o$ \\ \hline
$D_{eff}\left(10^{-12}m^2/s\right)$ & 4.639 & 4.644 & 3.513 & 3.593 & 2.709
& 2.862 \\ \hline
$D_{app}\left(10^{-12}m^2/s\right)$ & 4.634$\pm $0.002 & 4.646$\pm $0.008 & 
3.507$\pm $0.002 & 3.592$\pm $0.007 & 2.703$\pm $0.001 & 2.861$\pm $0.006 \\ 
\hline
\end{tabular}%
\end{center}
\caption{The values of $D_z$, $D_{eff}$ and $D_{app}$ for the simulated data
produced based on the mean static radius 50 nm and standard deviations 3, 15
and 25 nm at scattering angles 30$^\mathrm o$ and 90$^\mathrm o$,
respectively.}
\label{50Dall}
\end{table}

Table \ref{50Dall} shows that the value of $D_{eff}$ is consistent with that
of $D_{app}$ at any scattering angle investigated and the difference between
the values of $D_z$ and $D_{eff}$ is influenced by the scattering angle and
particle size distribution. For narrow particle size distributions, the
value of $D_z$ is consistent with that of $D_{eff}$ obtained at any
scattering angle investigated. For wide particle size distributions, the
difference between them is a function of the scattering angle. At a small
scattering angle, the result of $D_{eff}$ is a good approximate value of $%
D_z $.

The values of $\left( \frac{\sqrt{\mu _{2}}}{\langle \Gamma \rangle }\right)
_{z}$ and $\left( \frac{\sqrt{\mu _{2}}}{\langle \Gamma \rangle }\right)
_{eff}$ at scattering angles 30$^{\text{o}}$ and 90$^{\text{o}}$ for
different standard deviations 3, 5, 10, 15, 20 and 25 nm were calculated
using Eqs. \ref{Deffindex}, \ref{Dzindex} and the results of $\left( \frac{%
\sqrt{\mu _{2}}}{\langle \Gamma \rangle }\right) _{app}$ are obtained using
the first two moments method to fit the different simulated data,
respectively. All results of $\left( \frac{\sqrt{\mu _{2}}}{\langle \Gamma
\rangle }\right) _{z}$, $\left( \frac{\sqrt{\mu _{2}}}{\langle \Gamma
\rangle }\right) _{eff}$ and $\left( \frac{\sqrt{\mu _{2}}}{\langle \Gamma
\rangle }\right) _{app}$ for standard deviations 3, 15 and 25 nm are listed
in Table \ref{50indexall}.

\begin{table}[ht]
\begin{center}
\begin{tabular}{|c|c|c|c|c|c|c|}
\hline
$\sigma \left( nm\right) $ & \multicolumn{2}{|c}{3} & \multicolumn{2}{|c}{15}
& \multicolumn{2}{|c|}{25} \\ \hline
$\left( \sqrt{\mu _{2}}/\langle \Gamma\rangle \right) _{z}$ & 
\multicolumn{2}{|c}{0.059} & \multicolumn{2}{|c}{0.208} & 
\multicolumn{2}{|c|}{0.253} \\ \hline
$\theta $ & 30$^\mathrm o$ & 90$^\mathrm o$ & 30$^\mathrm o$ & 90$^\mathrm o$
& 30$^\mathrm o$ & 90$^\mathrm o$ \\ \hline
$\left( \sqrt{\mu _{2}}/\langle \Gamma\rangle \right)_{eff}$ & 0.059 & 0.059
& 0.208 & 0.209 & 0.253 & 0.255 \\ \hline
$\left( \sqrt{\mu _{2}}/\langle \Gamma\rangle \right) _{app}$ & 0.058$\pm $%
0.006 & 0.059$\pm $0.009 & 0.198$\pm $0.003 & 0.200$\pm $0.008 & 0.236$\pm $%
0.002 & 0.238$\pm $0.006 \\ \hline
\end{tabular}%
\end{center}
\caption{The values of $\left(\frac{\protect\sqrt{\protect\mu _{2}}}{\langle
\Gamma\rangle}\right)_z$, $\left(\frac{\protect\sqrt{\protect\mu _{2}}}{%
\langle \Gamma\rangle}\right)_{eff}$ and $\left(\frac{\protect\sqrt{\protect%
\mu _{2}}}{\langle \Gamma\rangle}\right)_{app}$ for the simulated data
produced based on the mean static radius 50 nm and standard deviations 3, 15
and 25 nm at scattering angles 30$^\mathrm o$ and 90$^\mathrm o$,
respectively.}
\label{50indexall}
\end{table}

Table \ref{50indexall} reveals that the value of $\left(\frac{\sqrt{\mu _{2}}%
}{\langle \Gamma\rangle}\right)_{eff}$ at any scattering angle investigated
is consistent with that of $\left(\frac{\sqrt{\mu _{2}}}{\langle
\Gamma\rangle}\right)_z$ and the difference between $\left(\frac{\sqrt{\mu
_{2}}}{\langle \Gamma\rangle}\right)_{eff}$ and $\left(\frac{\sqrt{\mu _{2}}%
}{\langle \Gamma\rangle}\right)_{app}$ at a scattering angle investigated is
affected by the particle size distribution. For narrow particle size
distributions, they are consistent. For wide particle size distributions,
the poly-dispersity index of apparent diffusion coefficient should be
measured at a $\tau $ range where $\tau_{max}$ changes to a small value in
order to obtain a good approximate result of the poly-dispersity index of
Z-average diffusion coefficient.

Next, the simulated data produced based on the mean static radius 120 nm and
standard deviations 8, 12, 24, 36, 48, 60 nm at scattering angles 30$^{\text{%
o}}$ and 90$^{\text{o}}$ were explored. The fit results for the simulated
data produced based on the standard deviations 12 nm, 60 nm at scattering
angles 30$^{\text{o}}$ and 90$^{\text{o}}$ are shown in Figs. 3 and 4,
respectively. The figures show that all the fit values obtained using the
first two moments method represent the simulated data very well,
respectively.

In order to obtain the results of $D_{eff}$ at scattering angles 30$^{\text{o%
}}$ and 90$^{\text{o}}$, and $D_{z}$, the mean static radius 120 nm and
standard deviations 8, 12, 24, 36, 48, 60 nm were input into Eqs. \ref{Deff}
and \ref{Dzff} to calculate and the results of $D_{app}$ are measured using
the first two moments method to fit the different simulated data,
respectively. All results of $D_{z}$, $D_{eff}$ and $D_{app}$ for standard
deviations 8, 36 and 60 nm are shown in Table \ref{120Dall}.

\begin{table}[ht]
\begin{center}
\begin{tabular}{|c|c|c|c|c|c|c|}
\hline
$\sigma \left( nm\right) $ & \multicolumn{2}{|c}{8} & \multicolumn{2}{|c}{36}
& \multicolumn{2}{|c|}{60} \\ \hline
$D_{z}\left(10^{-12}m^2/s\right)$ & \multicolumn{2}{|c}{1.765} & 
\multicolumn{2}{|c}{1.337} & \multicolumn{2}{|c|}{1.026} \\ \hline
$\theta $ & 30$^\mathrm o$ & 90$^\mathrm o$ & 30$^\mathrm o$ & 90$^\mathrm o$
& 30$^\mathrm o$ & 90$^\mathrm o$ \\ \hline
$D_{eff}\left(10^{-12}m^2/s\right)$ & 1.767 & 1.783 & 1.364 & 1.588 & 1.077
& 1.476 \\ \hline
$D_{app}\left(10^{-12}m^2/s\right)$ & 1.765$\pm $0.001 & 1.784$\pm $0.002 & 
1.362$\pm $0.001 & 1.587$\pm $0.002 & 1.075$\pm $0.001 & 1.476$\pm $0.002 \\ 
\hline
\end{tabular}%
\end{center}
\caption{The values of $D_z$, $D_{eff}$ and $D_{app}$ for the simulated data
produced based on the mean static radius 120 nm and standard deviations 8,
36 and 60 nm at scattering angles 30$^\mathrm o$ and 90$^\mathrm o$,
respectively.}
\label{120Dall}
\end{table}

Table \ref{120Dall} shows clearly the influences of particle size
distribution on the results of $D_{eff}$ and $D_{app}$. For narrow particle
size distributions, the values of $D_{eff}$ and $D_{app}$ obtained at any
scattering angle investigated are consistent with that of $D_z$. For wide
particle size distributions, the values of $D_{app}$ and $D_{eff}$ obtained
at a scattering angle investigated still are consistent and the difference
between $D_z$ and $D_{eff}$ depends on the scattering angle. In order to get
the good approximate value of $D_z$, the fit results of $D_{app}$ should be
obtained at a small scattering angle.

The values of $\left( \frac{\sqrt{\mu _{2}}}{\langle \Gamma \rangle }\right)
_{z}$ and $\left( \frac{\sqrt{\mu _{2}}}{\langle \Gamma \rangle }\right)
_{eff}$ at scattering angles 30$^{\text{o}}$ and 90$^{\text{o}}$ for
different standard deviations 8, 12, 24, 36, 48, 60 nm were calculated using
Eqs. \ref{Deffindex}, \ref{Dzindex} and the results of $\left( \frac{\sqrt{%
\mu _{2}}}{\langle \Gamma \rangle }\right) _{app}$ are obtained using the
first two moments method to fit the different simulated data, respectively.
All results listed in Table \ref{120indexall} show the value of $\left( 
\frac{\sqrt{\mu _{2}}}{\langle \Gamma \rangle }\right) _{eff}$ obtained at
any scattering angle is consistent with that of $\left( \frac{\sqrt{\mu _{2}}%
}{\langle \Gamma \rangle }\right) _{z}$ and the difference between $\left( 
\frac{\sqrt{\mu _{2}}}{\langle \Gamma \rangle }\right) _{eff}$ and $\left( 
\frac{\sqrt{\mu _{2}}}{\langle \Gamma \rangle }\right) _{app}$ is affected
by the particle size distribution. For narrow particle size distributions,
they are consistent. For wide particle size distributions, the
poly-dispersity index of apparent diffusion coefficient should be measured
at a $\tau $ range where $\tau _{max}$ changes to a small value in order to
obtain a good approximate result of the poly-dispersity index of Z-average
diffusion coefficient.

\begin{table}[ht]
\begin{center}
\begin{tabular}{|c|c|c|c|c|c|c|}
\hline
$\sigma \left( nm\right) $ & \multicolumn{2}{|c}{8} & \multicolumn{2}{|c}{36}
& \multicolumn{2}{|c|}{60} \\ \hline
$\left( \sqrt{\mu _{2}}/\langle \Gamma\rangle \right) _{z}$ & 
\multicolumn{2}{|c}{0.065} & \multicolumn{2}{|c}{0.208} & 
\multicolumn{2}{|c|}{0.253} \\ \hline
$\theta $ & 30$^\mathrm o$ & 90$^\mathrm o$ & 30$^\mathrm o$ & 90$^\mathrm o$
& 30$^\mathrm o$ & 90$^\mathrm o$ \\ \hline
$\left( \sqrt{\mu _{2}}/\langle \Gamma\rangle \right)_{eff}$ & 0.065 & 0.065
& 0.209 & 0.208 & 0.254 & 0.258 \\ \hline
$\left( \sqrt{\mu _{2}}/\langle \Gamma\rangle \right) _{app}$ & 0.065$\pm $%
0.005 & 0.065$\pm $0.008 & 0.199$\pm $0.002 & 0.196$\pm $0.003 & 0.237$\pm $%
0.002 & 0.246$\pm $0.003 \\ \hline
\end{tabular}%
\end{center}
\caption{The values of $\left(\frac{\protect\sqrt{\protect\mu _{2}}}{\langle
\Gamma\rangle}\right)_z$, $\left(\frac{\protect\sqrt{\protect\mu _{2}}}{%
\langle \Gamma\rangle}\right)_{eff}$ and $\left(\frac{\protect\sqrt{\protect%
\mu _{2}}}{\langle \Gamma\rangle}\right)_{app}$ for the simulated data
produced based on the mean static radius 120 nm and standard deviations 8,
36 and 60 nm at scattering angles 30$^\mathrm o$ and 90$^\mathrm o$,
respectively.}
\label{120indexall}
\end{table}

Finally, the situations for much larger particles were investigated. The
simulated data produced based on the mean static radius 260 nm and standard
deviations 16, 26, 52, 78, 104, 130 nm at a scattering angle of 30$^{\text{o}%
}$ were explored. The fit results for the simulated data produced based on
the standard deviations 26 nm, 130 nm at scattering angles 30$^{\text{o}}$
were shown in Fig. 5a and 5b, respectively. Figure 5 shows that all the fit
values obtained using the first two moments method are consistent with the
simulated data very well, respectively.

As above, using Eqs. \ref{Deff} and \ref{Dzff}, the values of $D_{z}$ and $%
D_{eff}$ at a scattering angle of 30$^{\text{o}}$ for different standard
deviations 16, 26, 52, 78, 104, 130 nm were obtained and using the first two
moments method to fit the different simulated data, the results of $D_{app}$
are got, respectively. All the results of $D_{z}$, $D_{eff}$ and $D_{app}$
for the standard deviations 16, 26, 78, 104 and 130 nm are listed in Table %
\ref{260Dall}.

\begin{table}[ht]
\begin{center}
\begin{tabular}{|c|c|c|c|c|c|}
\hline
$\sigma \left( nm\right) $ & 16 & 26 & 78 & 104 & 130 \\ \hline
$D_{z}\left(10^{-13}m^2/s\right)$ & 8.170 & 7.940 & 6.173 & 5.383 & 4.742 \\ 
\hline
$D_{eff}\left(10^{-13}m^2/s\right)$ & 8.212 & 8.047 & 6.835 & 6.366 & 6.039
\\ \hline
$D_{app}\left(10^{-13}m^2/s\right)$ & 8.203$\pm $0.004 & 8.037$\pm $0.004 & 
6.821$\pm $0.004 & 6.350$\pm $0.003 & 6.021$\pm $0.003 \\ \hline
\end{tabular}%
\end{center}
\caption{The values of $D_z$, $D_{eff}$ and $D_{app}$ for the simulated data
produced based on the mean static radius 260 nm and standard deviations 16,
26, 78, 104 and 130 nm at a scattering angle of 30$^\mathrm o$.}
\label{260Dall}
\end{table}

Table \ref{260Dall} reveals the same results as Tables \ref{50Dall} and \ref%
{120Dall}. For narrow particle size distributions, the values of $D_{eff}$
and $D_{app}$ obtained at a scattering angle of 30$^{\text{o}}$ are
consistent with that of $D_{z}$. For a wide particle size distribution, the
results of $D_{app}$ and $D_{eff}$ still are consistent and the values of $%
D_{z}$ and $D_{eff}$ can have a large difference. In order to get the good
approximate value of $D_{z}$, the DLS data for wide particle size
distributions should be measured at a much smaller scattering angle.

Using Eqs. \ref{Deffindex}, \ref{Dzindex}, the values of $\left( \frac{\sqrt{%
\mu _{2}}}{\langle \Gamma \rangle }\right) _{z}$ and $\left( \frac{\sqrt{\mu
_{2}}}{\langle \Gamma \rangle }\right) _{eff}$ at a scattering angle of 30$^{%
\text{o}}$ for different standard deviations 16, 26, 78, 104, 130 nm were
calculated and using the first two moments method to fit the different
simulated data, the results of $\left( \frac{\sqrt{\mu _{2}}}{\langle \Gamma
\rangle }\right) _{app}$ are obtained, respectively. All results listed in
Table \ref{260indexall} show the same results as Table \ref{50indexall} and %
\ref{120indexall}. The value of $\left( \frac{\sqrt{\mu _{2}}}{\langle
\Gamma \rangle }\right) _{eff}$ is consistent with that of $\left( \frac{%
\sqrt{\mu _{2}}}{\langle \Gamma \rangle }\right) _{z}$ and the difference
between $\left( \frac{\sqrt{\mu _{2}}}{\langle \Gamma \rangle }\right) _{eff}
$ and $\left( \frac{\sqrt{\mu _{2}}}{\langle \Gamma \rangle }\right) _{app}$
is affected by the particle size distribution. For narrow particle size
distributions, they are consistent.

\begin{table}[ht]
\begin{center}
\begin{tabular}{|c|c|c|c|c|c|}
\hline
$\sigma \left( nm\right) $ & 16 & 26 & 78 & 104 & 130 \\ \hline
$\left( \sqrt{\mu _{2}}/\Gamma \right) _{z}$ & 0.060 & 0.094 & 0.208 & 0.235
& 0.252 \\ \hline
$\left( \sqrt{\mu _{2}}/\Gamma \right) _{eff}$ & 0.060 & 0.095 & 0.211 & 
0.236 & 0.250 \\ \hline
$\left( \sqrt{\mu _{2}}/\Gamma \right) _{app}$ & 0.060$\pm $0.005 & 0.094$%
\pm $0.004 & 0.200$\pm $0.002 & 0.221$\pm $0.002 & 0.231$\pm $0.002 \\ \hline
\end{tabular}%
\end{center}
\caption{The values of $\left(\frac{\protect\sqrt{\protect\mu _{2}}}{\langle
\Gamma\rangle}\right)_z$, $\left(\frac{\protect\sqrt{\protect\mu _{2}}}{%
\langle \Gamma\rangle}\right)_{eff}$ and $\left(\frac{\protect\sqrt{\protect%
\mu _{2}}}{\langle \Gamma\rangle}\right)_{app}$ for the simulated data
produced based on the mean static radius 260 nm and standard deviations 16,
26, 78, 104 and 130 nm at a scattering angle of 30$^\mathrm o$.}
\label{260indexall}
\end{table}

In order to explore the influences of a wide particle size distribution on $%
D_{eff}$ for small particles in details, the values of $D_{eff}$ were
calculated using Eq. \ref{Deff} for $\langle R_{s}\rangle $ 30 nm and $%
\sigma $ 15 nm at different scattering angles. All results are shown in Fig.
6. Figure 6 reveals clearly that the DLS data should be measured at a small
scattering angle in order to obtain a good approximate value of $D_{z}$.

\section{CONCLUSION}

The values of the apparent and effective diffusion coefficients at a
scattering angle investigated are consistent and the difference between the
effective and Z-average diffusion coefficients is a function of the mean
particle size, size distribution and scattering angle. For the small
particles with narrow size distributions, the Z-average diffusion
coefficient can be got directly at any scattering angle. For the small
particles with wide size distributions, the Z-average diffusion coefficient
should be measured at a small scattering angle. For large particles, in
order to obtain a good approximate value of Z-average diffusion coefficient,
the wider the particle size distribution, the smaller the scattering angle
that the DLS data are measured. The poly-dispersity index of the effective
diffusion coefficient at a scattering angle investigated is consistent with
that of the Z-average diffusion coefficient and without considering the
influences of noises, the difference between the poly-dispersity indexes of
the Z-average and apparent diffusion coefficients is determined by the mean
particle size, size distribution and scattering angle together. For narrow
particle size distributions, they are consistent. For wide particle size
distributions, the poly-dispersity index of apparent diffusion coefficient
should be measured at a $\tau $ range where $\tau _{max}$ changes to a small
value in order to obtain a good approximate result of the poly-dispersity
index of Z-average diffusion coefficient.

Fig. 1 The fit results of $g^{\left( 2\right) }\left( \tau \right) $ for the
simulated data produced based on the mean static radius 50 nm and standard
deviation 5 nm. The circles show the simulated data and the line represents
the fit results obtained using the first two moments method. The results for
scattering angles 30$^{\text{o}}$ and 90$^{\text{o}}$ are shown in a and b,
respectively.

Fig. 2 The fit results of $g^{\left( 2\right) }\left( \tau \right) $ for the
simulated data produced based on the mean static radius 50 nm and standard
deviation 25 nm. The circles show the simulated data and the line represents
the fit results obtained using the first two moments method. The results for
scattering angles 30$^{\text{o}}$ and 90$^{\text{o}}$ are shown in a and b,
respectively.

Fig. 3 The fit results of $g^{\left( 2\right) }\left( \tau \right) $ for the
simulated data produced based on the mean static radius 120 nm and standard
deviation 12 nm. The circles show the simulated data and the line represents
the fit results obtained using the first two moments method. The results for
scattering angles 30$^{\text{o}}$ and 90$^{\text{o}}$ are shown in a and b,
respectively.

Fig. 4 The fit results of $g^{\left( 2\right) }\left( \tau \right) $ for the
simulated data produced based on the mean static radius 120 nm and standard
deviation 60 nm. The circles show the simulated data and the line represents
the fit results obtained using the first two moments method. The results for
scattering angles 30$^{\text{o}}$ and 90$^{\text{o}}$ are shown in a and b,
respectively.

Fig. 5 The fit results of $g^{\left( 2\right) }\left( \tau \right) $ for the
simulated data produced based on the mean static radius 260 nm and standard
deviations 26 and 130 nm. The circles show the simulated data and the line
represents the fit results obtained using the first two moments method. The
results for standard deviations 26 and 130 nm are shown in a and b,
respectively.

Fig. 6 The values of $D_{eff}$ for the simulated data produced based on the
mean static radius 30 nm and standard deviation 15 nm at different
scattering angles.

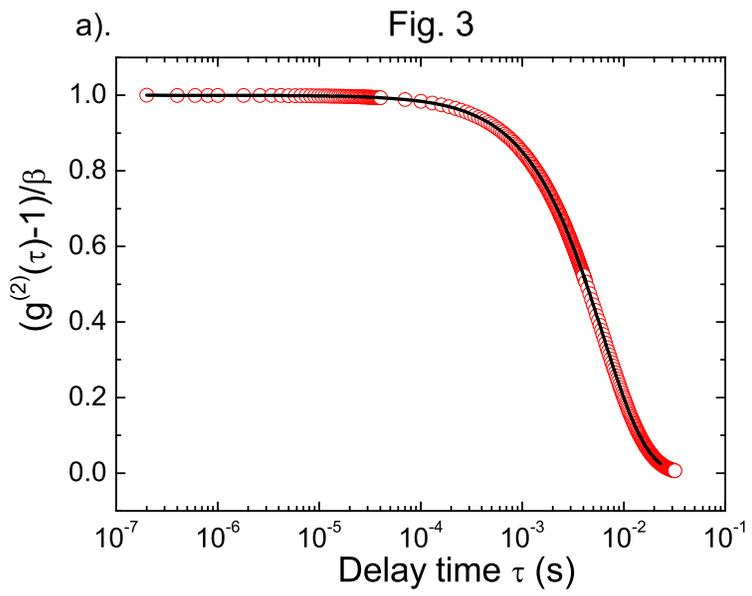
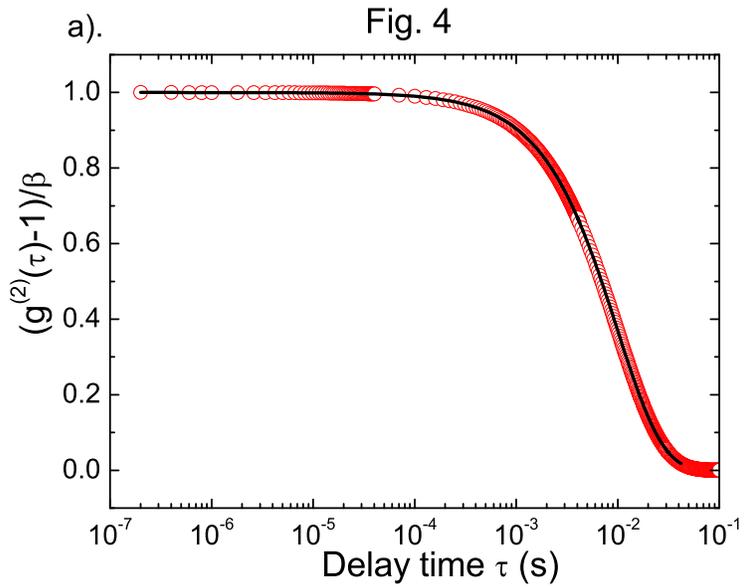
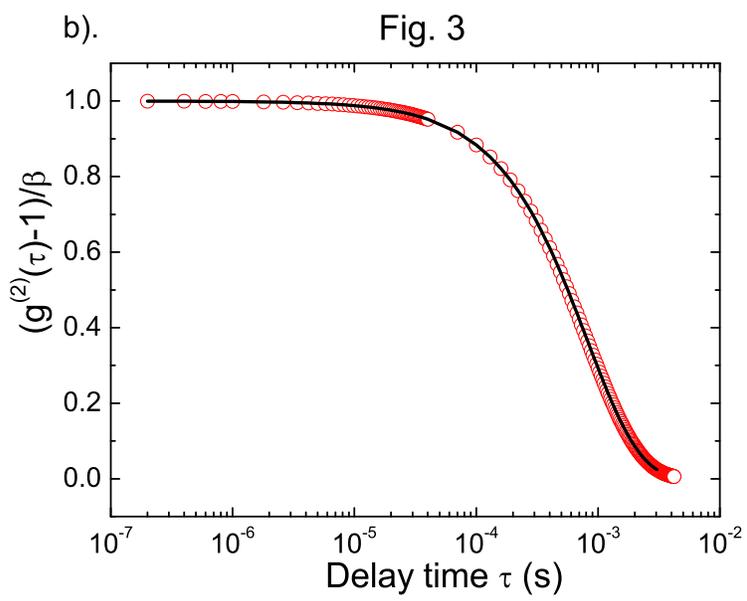
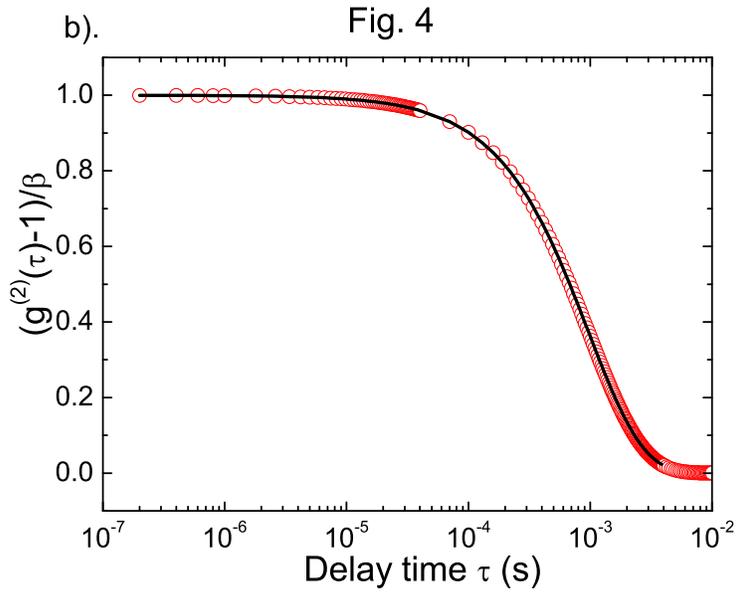
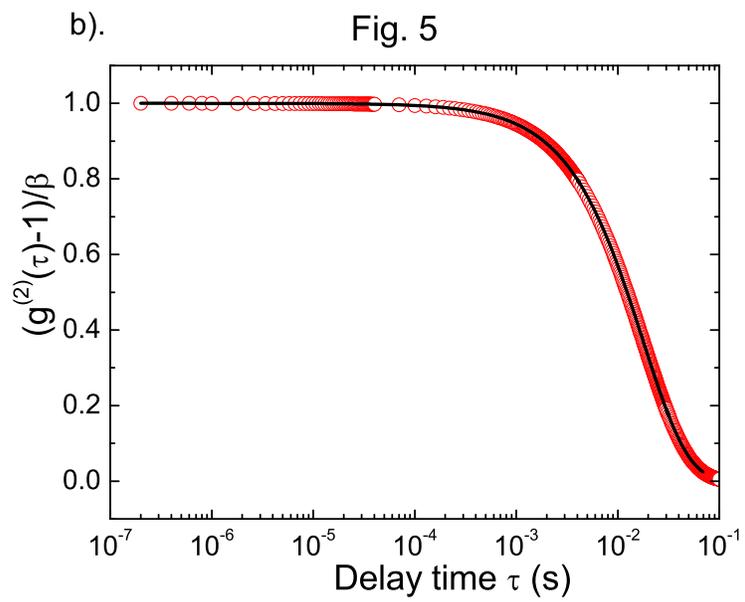
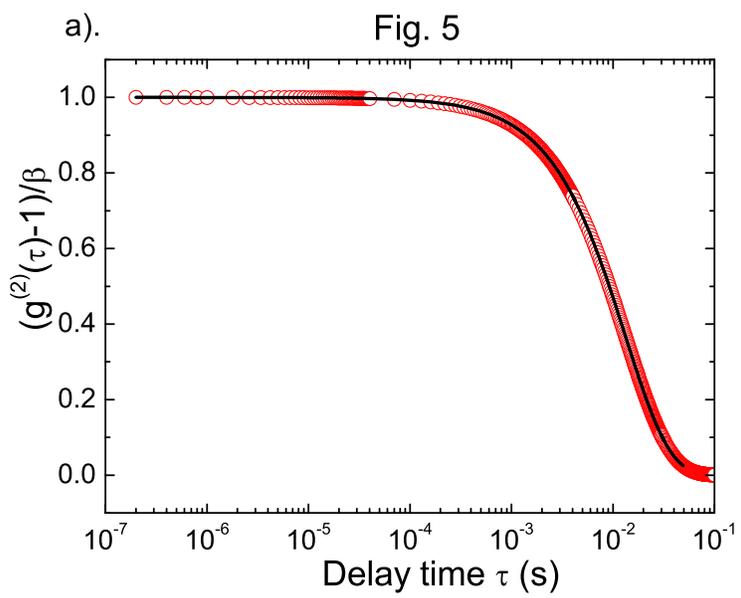
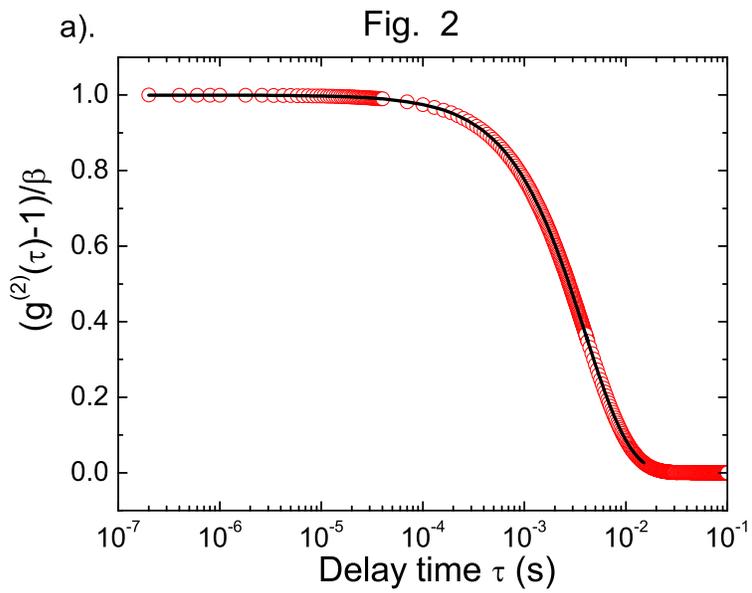
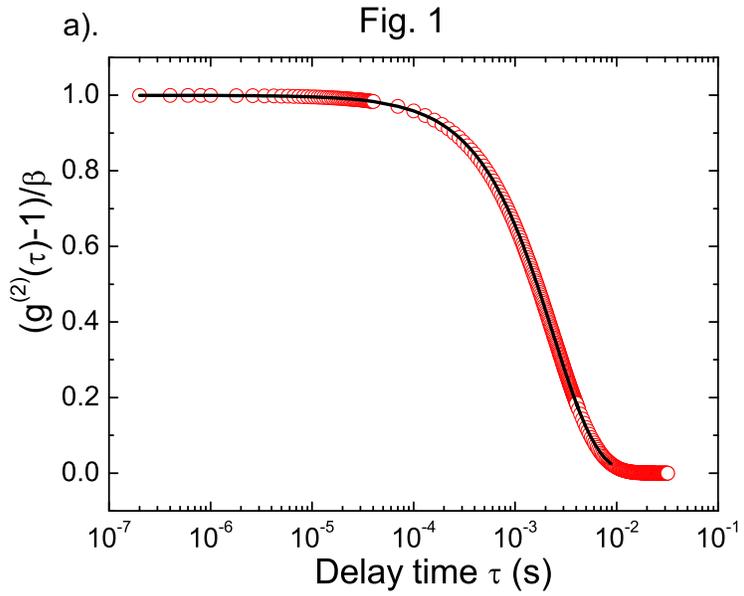
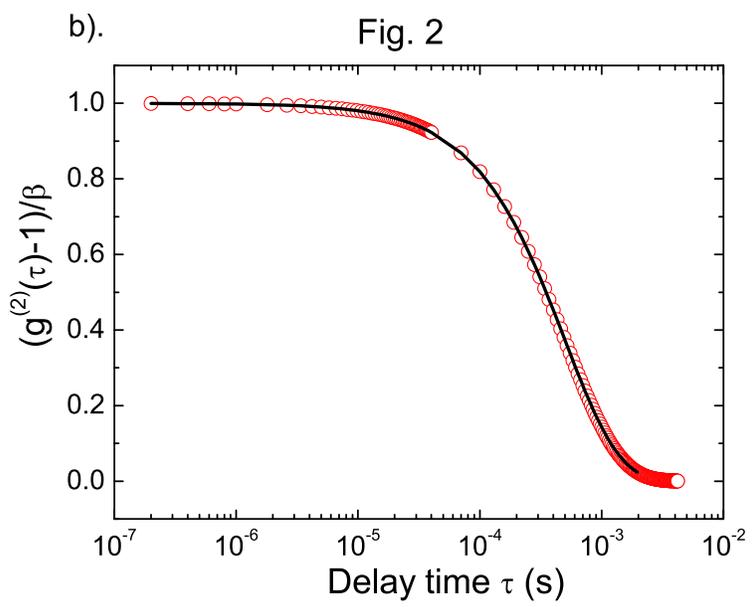
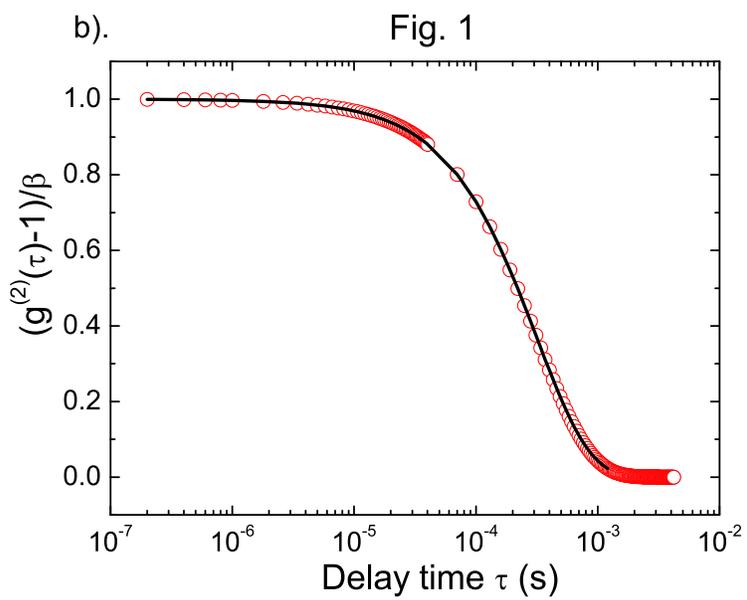
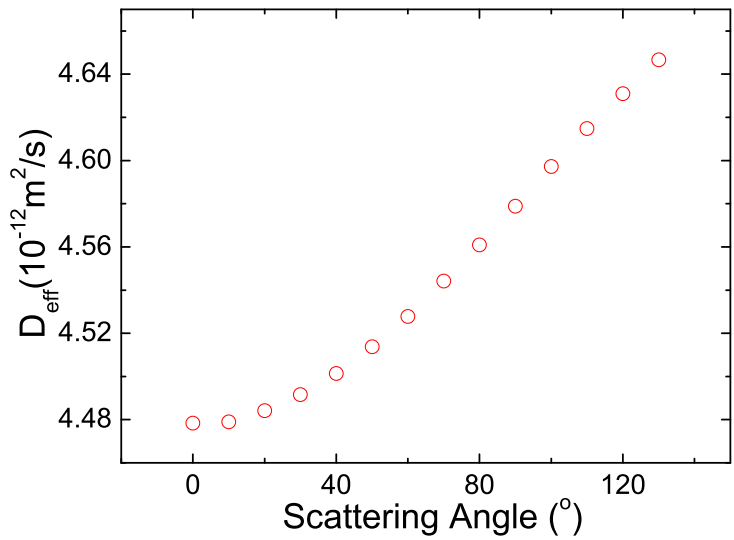
\end{document}